\documentclass
[prd,showpacs,preprintnumbers]{revtex4}
\def\lb{\label}
\def\be{\begin{equation}}
\def\ee{\end{equation}}
\def\ba{\begin{eqnarray}}
\def\ea{\end{eqnarray}}
\def\pd{\partial}
\def\ol{\overline}
\def\z{\zeta}
\def\bz{\overline\zeta}
\def\bZ{\overline{Z}}
\newcommand{\e}{{\rm e}}
\begin{document}

\title{
   \begin{flushright} \begin{small}
     LAPTH-1101/05, DTP-MSU/05-09,   \\
  \end{small} \end{flushright}
\vspace{.5cm}
Comment on ``What does the Letelier-Gal'tsov metric describe?''}

\author
{G. Cl\'ement$^{a}$\thanks{email: gclement@lapp.in2p3.fr} , D.V.
Gal'tsov$^{a,b}$\thanks{email: galtsov@mail.phys.msu.ru} and P.S.
Letelier$^{c}$\thanks{email: letelier@ime.unicamp.br}}
\affiliation {  $^{a}$Laboratoire de Physique Th\'eorique LAPTH
(CNRS), B.P.110, F-74941
Annecy-le-Vieux cedex, France \\
 $^{b}$Department of Theoretical
Physics, Moscow State University, 119899, Moscow, Russia \\
 $^{c}$Departamento de Matem\'atica Aplicada-IMECC, Universidade
Estadual de Campinas,\\ 13083-970 Campinas, S.P., Brazil}

\begin{abstract}
We show that the Letelier-Gal'tsov (LG) metric describing multiple 
crossed strings in relative motion \cite{lg} does solve the Einstein 
equations, in spite of the discontinuity uncovered recently by 
Krasnikov \cite{krasn}, provided the strings are straight and moving 
with constant velocities.
\end{abstract}
\maketitle

In a recent note \cite{krasn}, Krasnikov raised an interesting
question about the continuity of the LG metric. He showed, on the
example of a (parallel) two-cosmic string  LG  metric, that the
metric component $g_{ty}$ (where $y$ is the coordinate orthogonal
to strings and their relative displacement) is generically
discontinuous across the cuts associated with the positions of
strings. Arguing that the metric must be continuous, he concluded
that the strings should be at rest with respect to each other.

Before proceeding with a more elaborate analysis, we note it would
be too strong to require the metric to be everywhere continuous.
To define a spacetime, one needs a set of overlapping,
diffeomorphic charts. We will show that, provided the string
motion is geodesic and the strings do not collide, the Riemann tensor
computed from the discontinuous metric vanishes outside the strings. 
It follows that there is a set of overlapping charts such that the 
metric is continuous in each.

Let us first briefly reformulate Krasnikov's argument. For simplicity,
we shall consider here only the case of moving parallel strings, which 
can be reduced to that of moving conical singularities in 2+1
dimensions. The reduced LG metric can be written in the ADM form, 
\be ds^2 = - N^2dt^2
+ h_{ij}(dx^i+N^idt)(dx^j+N^jdt), \ee with \be\lb{lg} N = 1,\quad
h_{\z\bz} = \frac12\bZ_{,\bz}Z_{,\z}, \quad
N_{\z}=h_{\z\bz}N^{\bz}= \frac12\bZ_{,t}Z_{,\z} \ee ($\z = x+iy$),
where, in the case of two strings,
\be Z(t,\z) = \int_{\z_0}^{\z}\psi(t,\xi)\,d\xi\,, \qquad
\psi(t,\xi) = (\xi-\alpha_1(t))^{\mu_1}(\xi-\alpha_2(t))^{\mu_2}
\ee 
($\mu_i >- 1$). This is defined only in the cut complex plane with
two line cuts extending from the two conical singularities to infinity. 
We choose as the first cut the horizontal half-axis $[\alpha_1, 
\alpha_1 +\infty[$. The second cut, starting in $\alpha_2$, will be 
assumed not to intersect the first cut (in this choice we differ from 
\cite{krasn}). Following \cite{krasn}, we also choose a gauge such that
$\alpha_1(t) = 0$, and  $\alpha_2(t)$ and $\z_0$ real, with $\alpha_2 <
\z_0 < 0$. 

We choose $\z$ to be on the first cut, and wish to
evaluate at a given time $t$ the discontinuity of 
\be N_{\z} =
\frac12\psi(\z)\ol{Z_{,t}(\z)}. \ee 
First, 
\be\lb{Npm}
N_{\z}(\z\pm i0)=
\frac12\psi(\z)\e^{i\pi\mu_1(1\mp1)}\ol{Z_{,t}(\z\pm i0)}, \ee
with 
\be Z_{,t}(\z\pm i0) = 
-\mu_2\dot{\alpha}_2\int_{\Gamma_{\pm}}\frac{\psi(\xi)}
{\xi-\alpha_2}\,d\xi\,. \ee 
Choosing $\Gamma_+$ ($\Gamma_-$) to go
first along an arbitrary path from $\z_0$ (a fixed point anywhere
in the $\z$ plane) to $\alpha_1=0$, then (after a small upper
(lower) half-circle around the origin) to follow the upper (lower)
bank $\gamma_+$ ($\gamma_-$) of the cut from $0$ to $\z$,
we obtain  
\be\lb{Npm1} N_{\z}(\z\pm i0)=
\frac12\left(\ol{\chi(\z)}\psi(\z)+ V\e^{\mp i\pi\mu_1}
\psi(\z)\right)\,, \ee 
where 
\be V(t) =
e^{-i\pi\mu_1}\dot{Z}(0) 
\ee 
is the velocity of the first string (real in our gauge), and 
\be\lb{chi} \quad \chi(\z) =
-\dot{\alpha}_2\int_{0}^{\z}
\frac{\psi(\xi)}{\xi-\alpha_2}\,d\xi\,. \ee

Extending (\ref{Npm1}) to $y = {\rm Im}(\z) \neq
0$, we obtain, in the vicinity of the cut, \ba\lb{Nz} N_{\z}(\z)
&=& \frac12\left(\ol{\chi(\z)}\psi(\z)+
V\,\e^{-i\pi\mu_1\epsilon(y)}
\psi(\z)\right) \nonumber \\
&=& \frac12\left(\ol{\chi(\z)}+V\cos(\pi\mu_1)\right)\psi(\z)-
\frac{i}2\beta\psi(\z)\epsilon(y)\,, \ea where $\epsilon(y)$ is
the sign function, and $\beta = V\sin(\pi\mu_1)$.
So, while the lapse $N$ and the two-metric $h_{ij}$ are by
construction single-valued and thus continuous, the shift $N_{\z}$
is not continuous across the cut, except in the static case
$V(t) = 0$. A priori this
discontinuity will lead to delta and gradient of delta
contributions to the Einstein tensor. We shall see that the
gradient of delta contribution vanishes identically. However the
delta contribution remains, so that generically there must be a
matter source along the cut. We shall now show that this delta
contribution vanishes, iff the motion is geodesic, $\dot{\beta}(t)
= 0$. The Riemann tensor (completely determined in 2+1 dimensions 
by the Einstein tensor) then vanishes outside the point sources $\z =
\alpha_i(t)$.  

For this purpose we use the ($1+2$)-dimensional vacuum Einstein
equations written in the ADM form  \ba {\cal H} &\equiv&
-h^{1/2}R(h) +
h^{-1/2}\left(\pi^{ij}\pi_{ij}-\pi^2\right) = 0\,,  \lb{H}\\
{\cal H}^i &\equiv& -2{\pi^{ij}}_{;j} = 0 \lb{Hi} \ea (constraint
equations), and \ba \pd_t h_{ij} &=&
2h^{-1/2}\left(\pi_{ij}-h_{ij}\pi\right) + N_{i;j} +
N_{j;i}\,, \lb{K} \\
\pd_t \pi^{ij} &=&
\frac12h^{-1/2}\left(h^{ij}\left(\pi^{kl}\pi_{kl}-\pi^2\right) -
4\left(\pi^{ik}{\pi^j}_k - \pi^{ij}\pi\right)\right) \nonumber \\
&& + (\pi^{ij}N^k)_{;k} - {N^i}_{;k}\pi^{kj} - {N^j}_{;k}\pi^{ki}
\lb{evol} \ea (evolution equations), where $h_{ij}$ is the
two-dimensional metric, $N=1$, $\pi = {\pi^k}_k$, and 
$\pi^{ij}$ are the momenta conjugate to the $h_{ij}$, related to 
the extrinsic curvature by 
\be \pi_{ij} = h^{1/2} \left(K
h_{ij}-K_{ij}\right).\ee 
Using the result (\ref{Nz}) and the non-vanishing Christoffel symbols
$\Gamma^{\z}_{\z\z} = \psi_{,\z}/{\psi}$
(together with its complex conjugate), we obtain in terms of the
real spatial coordinates $x, y$  
\be\lb{pi} \pi^{xx} = \beta\psi^{-1}(x)\delta(y)\,, \quad
\pi^{xy} = 0\,, \quad \pi^{yy} = 0\,. \ee 
The component $\pi^{xx}$
vanishes only in the static case ($\beta = 0$), reflecting the
fact that in this case the two-dimensional metric (\ref{lg}) is flat 
outside the conical singularities $\zeta = \alpha_i$.

Inserting this result into the constraint equations, we find that
the Hamiltonian constraint (\ref{H}) is identically satisfied
(using the fact that $R(h) = 0$ for the metric (\ref{lg})), while
the momentum constraint (\ref{Hi}) is satisfied on account of
$\pd_y(|\psi|^2)|_{y = 0} = 0$. The evolution equations (\ref{K})
have already been used to compute the momenta $\pi^{ij}$. There
remain only the evolution equations (\ref{evol}), which may be
rewritten \be \pd_t \pi^{ij} = (\pi^{ij}N^k)_{,k} -
{N^i}_{,k}\pi^{kj} - {N^j}_{,k}\pi^{ki} \lb{evol1}\,. \ee Both
sides of the ($yy$) equation vanish identically. The ($xy$)
equation is satisfied on account of $\pd_xN_y|_{y = 0} = 0$.
There only remains the potentially dangerous ($xx$) equation. 
For its left-hand side we obtain  
\be\lb{lh} \pd_t\pi^{xx} = \left(\dot{\beta} +
\frac{\beta\mu_2\dot\alpha_2}{x-\alpha_2}\right) \psi^{-1}(x)
\delta(y)\,. \ee 
For the right-hand side, using
\be
\pd_xN^x = -\frac{\mu_2\dot\alpha_2}{x-\alpha_2} 
 -\psi^{-1}\psi_{,x}N^x \quad \mbox{for} \quad y = 0 \,,
\ee  
we find 
\be\lb{rh}
N^x\pd_x\pi^{xx} - \pi^{xx}\pd_xN^x =
\frac{\beta\mu_2\dot\alpha_2}{x-\alpha_2}\psi^{-1}(x)\delta(y)\,.
\ee 
Comparison of (\ref{lh}) and (\ref{rh}) shows that the
($xx$) component of the evolution equation (\ref{evol1}) is
satisfied iff 
\be \dot\beta = 0\,. \ee 

The argument can be generalized to show that time derivatives 
\be \dot{Z}[t,z,\z=\alpha_i(t,z)]=Z_{,t}(\alpha_i) + Z_{,\z}(\alpha_i)
\dot{\alpha_i} = {\rm const.}\ee 
for each string, i.e. the strings must move with constant velocities. 
The extension to the case of non-parallel strings is straightforward
and leads to the conclusion that the derivatives with respect to $z$ 
must satisfy
\be Z'[t,z,\z=\alpha_i(t,z)]={\rm const.},\ee 
meaning that the strings are straight. Therefore the complex trajectories
$Z[\alpha_i]$ must be linear functions of $t$ and $z$, or, in
geometric terms, the world-sheets of the strings must be totally
geodesic submanifolds. We have shown here that this well-known
property of self-gravitating cosmic strings \cite{vickers,UHIM} is the
necessary and sufficient condition for the LG metric to represent a
system of crossed straight cosmic strings moving in otherwise empty spacetime.

D.G. thanks LAPTH (Annecy) for hospitality while this note was
being written. His work was also supported in part by the RFBR
grant 02-04-16949.

\end{document}